# Capacity Expansion of High Renewable Penetrated Energy Systems Considering Concentrating Solar Power for Seasonal Energy Balance

*Jing Li[a], Tianguang Lu[a,b]\*, Xinning Yi[a], Shaorui Wang[a], Xueqian He[a]*

[a]*School of Electrical Engineering, Shandong University, Jinan 250061, China*
[b]*School of Engineering and Applied Sciences and Harvard China Project, Harvard University, Cambridge, MA 02138, United States*

ABSTRACT

With the increasing proportion of variable renewable energy which owns fluctuation characteristics and the promotion of the "Clean Heating" policy, the seasonal energy imbalance of the system has been more and more challenging. There is a lack of effective means to mitigate this challenge under the background of gradual compression of the traditional thermal unit construction. Concentrating solar power (CSP) is a promising technology to replace thermal units by integrating emergency boilers to cope with extreme weather, and can meet long-time energy balance as a seasonal peak regulation source. In this paper, we propose a long-term high-resolution expansion planning model of the energy system under high renewable penetration which integrates CSP technology for seasonal energy balance. With the projection to 2050, by taking the energy system in Xinjiang province which is a typical area of the "Clean Heating" project with rich irradiance as a case study, it shows that the optimal deployment of CSP and electric boiler (EB) can reduce the cost, peak-valley difference of net load and renewable curtailment by 8.73%, 19.72% and 58.24% respectively at 65% renewable penetration compared to the base scenario.

## 1. Introduction

Under the challenge of energy crisis and climate change, high penetration of renewable energy will be the feature of energy systems in the future [1]. According to International Renewable Energy Agency data, the newly built capacity of renewable energy generators is more than 295,000 MW, accounting for 83% of total newly built capacity of generators all over the world in 2022. By the end of 2022, the total capacity of global renewable energy generation reaches 3,372,000 MW. More than 130 nations and regions have established climate goals to reduce carbon emissions. [2]. Especially in China, with the implementation of the carbon peaking and carbon neutrality strategy, the scale of renewable energy continues to expand, and the construction space of traditional thermal units will be further compressed, while the energy demand is still growing steadily. In 2022, the newly built capacity of renewable energy generators is 152,000 MW, accounting for 76.2% of the newly built capacity of all kinds of generators in China. By the end of 2022, the total capacity of renewable energy generators reached 1,213,000 MW, accounting for 47.3% of total capacity of all kinds of generators, which is 2.5% higher than 2021. Among them, the total capacity of wind power generators is 365,000 MW and the total capacity of solar power generators is 393,000 MW. Renewable energy will become the main body of installed capacity of generators in China [3]. As a global leader in renewable energy capacity growth, it makes positive contributions to global carbon emissions reduction.

However, the uncertainty of renewable energy makes it difficult for the system to achieve full-time energy balance. The seasonal imbalance of renewable energy generation is an inevitable problem when the proportion of renewable energy increases to a certain stage [4]. The proportion curve

---

\* *Corresponding author.*
 E-mail address: tlu@sdu.edu.cn (T. Lu).

**Nomenclature**

*Acronyms and abbreviations*

| | | | |
|---|---|---|---|
| CSP | Concentrating solar power | $\Delta t$ | Duration of a time interval |
| SF | Solar field | $a_w$ | Newly installed cost of wind units |
| TES | Thermal energy storage | $f_w$ | Fixed operational maintenance cost of wind units |
| PB | Power block | $a_s$ | Newly installed cost of PV units |
| HTF | Heat-transfer fluid | $f_s$ | Fixed operational maintenance cost of PV units |
| EB | Electric boiler | $J$ | Number of groups of CSP units |
| PV | Photovoltaic | $a_{CSP,j}$ | Newly installed cost of CSP units |
| CHP | Combined heat and power | $f_{CSP,j}$ | Fixed operational maintenance cost of CSP units |
| LCOE | Levelized cost of energy | $H$ | Number of groups of CHP units |

*Parameters*

| | | | |
|---|---|---|---|
| | | $a_{CHP,h}$ | Newly installed cost of CHP units |
| $\eta_{SF}$ | Efficiency factor of SF | $f_{CHP,h}$ | Fixed operational maintenance cost of CHP units |
| $DNI$ | Value of solar direct normal irradiance | $c_{CHP,h}$ | Fuel costs of CHP units |
| $\gamma$ | Heat dissipation rate of TES | $st_{CHP,h}$ | Start-up costs of CHP units |
| $\eta_{TES}^{cha}$ | Charging efficiency factor of TES | $c_{v,h}$ | Parameters of the operational domain of CHP units |
| $\eta_{TES}^{dis}$ | Discharging efficiency factor of TES | $a_{EB}$ | Newly installed cost of EB |
| $\eta_{PB}$ | Efficiency factor of PB | $f_{EB}$ | Fixed operational maintenance cost of EB |
| $Q_{j,\min}^{CSP}$ | Minimum value of state of charge (SOC) of TES | $c_{EB}$ | Operational cost of EB |
| $Q_{j,\max}^{CSP}$ | Maximum value of state of charge (SOC) of TES | $c_c$ | Punishment cost factor due to renewable curtailment |
| $\underline{\alpha}_{i,t}^{CSP}$ | Ratio of minimum power output of CSP unit $i$ to the nameplate capacity | $r$ | Discount ratio of investment |
| $\overline{\alpha}_{i,t}^{CSP}$ | Ratio of maximum power output of CSP unit $i$ to the nameplate capacity | $\overline{\mu}_{coal,m}$ | Ratio of maximum power output of coal unit group $m$ to the online capacity |
| $P_{i,N}^{CSP}$ | Nameplate capacity of CSP unit $i$ | $\overline{\mu}_{CHP,h}$ | Ratio of maximum power output of CHP unit group $h$ to the online capacity |
| $P_{i,\min}^{CSP}$ | Minimum power output of CSP unit $i$ | $\alpha_t$ | Hourly capacity factors for wind units |
| $P_{i,\max}^{CSP}$ | Maximum power output of CSP unit $i$ | $\beta_t$ | Hourly capacity factors for PV units |
| $I$ | Number of CSP units within the group | $\lambda_t$ | Hourly capacity factors for CSP units |
| $\underline{A}_{j,t}^{CSP}$ | Ratio of minimum power output of CSP unit group $j$ to the online capacity at time $t$ | $R_t^d$ | Reserve requirement related to power demand |
| | | $R_w$ | Forecasting errors for wind units |
| $\overline{A}_{j,t}^{CSP}$ | Ratio of maximum power output of CSP unit group $j$ to the online capacity at time $t$ | $R_s$ | Forecasting errors for PV units |
| | | $R_c$ | Forecasting errors for CSP units |
| $P_{j,\min}^{CSP}$ | Minimum electrical power output of CSP unit group $j$ | $D_{E,t}$ | Electrical power demand |
| $P_{j,\max}^{CSP}$ | Maximum electrical power output of CSP unit group $j$ | $p$ | Specified renewable penetration rate to meet the renewable portfolio standards (RPS) |
| $R_{CSP,j}^U$ | Upward ramping coefficient of CSP unit group $j$ | $D_{H,t}$ | Heat demand (Thermal energy demand) |
| $R_{CSP,j}^D$ | Downward ramping coefficient of CSP unit group $j$ | $Q_{h,\min}^{CHP}$ | Minimum thermal energy output of CHP unit group $h$ |
| $T_{CSP,j}^U$ | Minimum online time | $Q_{h,\max}^{CHP}$ | Maximum thermal energy output of CHP unit group $h$ |
| $T_{CSP,j}^D$ | Minimum offline time | $P_{h,\min}^{CHP}$ | Minimum electrical power output of CHP unit group $h$ |
| $T$ | Time periods | $P_{h,\max}^{CHP}$ | Maximum electrical power output of CHP unit group $h$ |
| $M$ | Number of groups of coal units | $\underline{A}_{h,t}^{CHP}$ | Ratio of minimum power output of CHP unit group $h$ to the online capacity at time $t$ |
| $a_{coal,m}$ | Newly installed cost of coal units | | |
| $f_{coal,m}$ | Fixed operational maintenance cost of coal units | $\overline{A}_{h,t}^{CHP}$ | Ratio of maximum power output of CHP unit group $h$ to the online capacity at time $t$ |
| $c_{coal,m}$ | Fuel costs of coal units | | |
| $st_{coal,m}$ | Start-up costs of coal units | $c_{m,h}$ | Parameters of the operational domain of CHP units |

| Symbol | Description | Symbol | Description |
|---|---|---|---|
| $R_{CHP,h}^{U}$ | Upward ramping coefficient of CHP unit group $h$ | $Q_{j,t}^{TES-HD}$ | Thermal energy output to heat demand from TES |
| $R_{CHP,h}^{D}$ | Downward ramping coefficient of CHP unit group $h$ | $C_{coal}$ | Cost of traditional coal units |
| $I'$ | Number of CHP units within the group | $C_w$ | Cost of wind units |
| $\eta_{EB}$ | Efficiency factor of EB | $C_s$ | Cost of PV units |
| $Q_{\min}^{EB}$ | Minimum thermal energy output of EB | $C_{CSP}$ | Cost of CSP units |
| $Q_{\max}^{EB}$ | Maximum thermal energy output of EB | $C_{CHP}$ | Cost of CHP units |
| $\omega_1$ | Carbon emission factor of coal units | $C_{EB}$ | Cost of EB units |
| $\omega_2$ | Carbon emission factor of CHP units | $C_c$ | Punishment cost caused by wind and solar curtailment |
| $I_{coal,m}^{0}$ | Existing capacity of coal unit group $m$ | $I_{coal,m}$ | Newly installed capacity of coal unit group $m$ |
| $I_{w}^{0}$ | Existing capacity of wind units | $\bar{I}_{coal,m}$ | Total capacity of coal unit group $m$ |
| $I_{s}^{0}$ | Existing capacity of PV units | $P_{m,t}^{coal}$ | Power output of coal unit group $m$ at time $t$ |
| $I_{CSP,j}^{0}$ | Existing capacity of CSP unit group $j$ | $S_{coal,m,t}^{O}$ | Online capacity of coal unit group $m$ at time $t$ |
| $I_{CHP,h}^{0}$ | Existing capacity of CHP unit group $h$ | $S_{coal,m,t}^{U}$ | Start-up capacity of coal unit group $m$ at time $t$ |
| $I_{EB}^{0}$ | Existing capacity of EB | $S_{coal,m,t}^{D}$ | Shut-down capacity of coal unit group $m$ at time $t$ |
| **Variables** | | $I_w$ | Newly installed capacity of wind units |
| Binary variables | | $\bar{I}_w$ | Total capacity of wind units |
| $x_{i,t}^{CSP}$ | On/off status of unit $i$ at time $t$ ($I_{i,t}^{CSP}=1$ indicate unit $i$ is online at time $t$) | $P_t^w$ | Power output of wind units at time $t$ |
| | | $P_{t,\max}^{w}$ | Maximum power output of wind units at time $t$ |
| $u_{i,t}^{CSP}$ | Startup behavior of unit $i$ at time $t$ ($u_{i,t}^{CSP}=1$ indicate unit $i$ starts at time $t$) | $I_s$ | Newly installed capacity of PV units |
| | | $\bar{I}_s$ | Total capacity of PV units |
| $d_{i,t}^{CSP}$ | Shutdown behavior of unit $i$ at time $t$ ($d_{i,t}^{CSP}=1$ indicate unit $i$ is shut down at time $t$) | $P_t^s$ | Power output of PV units at time $t$ |
| | | $P_{t,\max}^{s}$ | Maximum power output of PV units at time $t$ |
| Integer variables | | $S_{SF}$ | Area of mirrors in SF |
| $\hat{S}_{CSP,j,t}^{O}$ | The sum of nameplate capacities of units within group $j$ that are operating at time $t$ | $I_{CSP,j}$ | Newly installed capacity of CSP unit group $j$ |
| | | $\bar{I}_{CSP,j}$ | Total capacity of CSP unit group $j$ |
| $\hat{S}_{CSP,j,t}^{U}$ | The sum of nameplate capacities of units within group $j$ starting up at time $t$ | $I_{CHP,h}$ | Newly installed capacity of CHP unit group $h$ |
| | | $\bar{I}_{CHP,h}$ | Total capacity of CHP unit group $h$ |
| $\hat{S}_{CSP,j,t}^{D}$ | The sum of nameplate capacities of units within group $j$ shutting down at time $t$ | $P_{h,t}^{CHP}$ | Electrical power output of CHP unit group $h$ at time $t$ |
| | | $Q_{h,t}^{CHP}$ | Thermal energy output of CHP unit group $h$ at time $t$ |
| Continuous Variables | | $S_{CHP,h,t}^{O}$ | Online capacity of CHP unit group $h$ at time $t$ |
| $S_{CSP,j,t}^{O}$ | Online capacity of CSP unit group $j$ at time $t$ | $S_{CHP,h,t}^{U}$ | Start-up capacity of CHP unit group $h$ at time $t$ |
| $S_{CSP,j,t}^{U}$ | Start-up capacity of CSP unit group $j$ at time $t$ | $S_{CHP,h,t}^{D}$ | Shut-down capacity of CHP unit group $h$ at time $t$ |
| $S_{CSP,j,t}^{D}$ | Shut-down capacity of CSP unit group $j$ at time $t$ | $S_{CHP,h}$ | Total capacity of CHP unit group $h$ |
| $S_{CSP,j}$ | Total capacity of CSP unit group $j$ | $I_{EB}$ | Newly installed capacity of EB |
| $P_{j,t}^{CSP}$ | Electrical power output of CSP unit group $j$ at time $t$ | $\bar{I}_{EB}$ | Total capacity of EB |
| $Q_{j,t}^{SF-HTF}$ | Thermal energy converted input to HTF by SF | $P_t^{EB}$ | Electrical power input of EB at time $t$ |
| $Q_{j,t}^{TES-HTF}$ | Thermal energy converted input to HTF by TES | $Q_t^{EB}$ | Thermal energy output of EB at time $t$ |
| $Q_{j,t}^{HTF-TES}$ | Thermal energy converted input to TES by HTF | $Q_t^{EB-TES}$ | Thermal energy converted input to TES by EB |
| $Q_{j,t}^{HTF-PB}$ | Thermal energy converted input to PB by HTF | $\eta_{cur}$ | Renewable energy curtailment rate |
| $Q_{j,t}^{cur}$ | Energy loss in SF | $E_{CO_2,t}$ | Carbon emission |
| $Q_{j,t}^{CSP}$ | State of charge (SOC) of TES at time $t$ | $C_{inve,t}$ | Investment cost |
| $Q_{j,t}^{TES,cha}$ | Charging energy of TES | $C_{main,t}$ | Maintenance cost |
| $Q_{j,t}^{TES,dis}$ | Discharging energy of TES | $C_{oper,t}$ | Operation cost |
| $Q_{n,t}^{CHP,cur}$ | Thermal energy input to TES by CHP plants | $E_t$ | Electrical power output |
| $Q_{j,t}^{EB-TES}$ | Thermal energy converted input to TES by EB | $H_t$ | Thermal energy output |

of monthly wind and PV power output and load for the whole year of a province in China is shown in Fig. 1. As shown in Fig. 1(a), we can find that the power demand is at its peak in summer, while the wind power output peaks in spring and winter. As shown in Fig. 1(b), we can find that the power demand is also a bit high in winter due to "Clean Heating" policy [5] such as "Electric Heating" policy, while the PV power output is very low in winter. Thus, there is an obvious seasonal mismatch between renewable power supply and power demand for the whole year. As the rate of renewable penetration rises, the seasonal imbalance will be more prominent and the peak regulation demand of energy systems will also change greatly. Compared to the traditional energy systems which mainly include traditional thermal units, the peak regulation demand of the high renewable penetrated energy systems is not only caused by load fluctuations, but also related to renewable energy usage. It faces the demand of both upward and downward peak regulation, and the peak-valley gap of the system is widening. Due to the imbalance of seasonal resources and the lack of flexible resources, the energy system will face a serious situation of unstable power supply. In 2016, there is a large number of unplanned customer outages in South Australia resulting in a loss of 300,000 kW of load when the rate of renewable energy penetration exceeds 30%, which is caused by lower-than-forecast wind power output and a serious lack of flexible resource [6]. In 2020, there is a power shortage for over 400,000 customers for about 1 hour in California because of the significant decrease in photovoltaic and wind power output due to the weather. The total capacity of photovoltaic units is more than 27.4 GW in California, but the renewable power output is only 3.3 GW at that time [7]. Currently, the rate of variable renewable curtailment reached as high as 20% in northwest area of China during the winter [8]. These events show that variable renewable energy cannot completely replace conventional stable generators, and it is difficult to cope with the power supply gap caused by extreme weather. There is an urgent need to find a regulating source that is both clean and stable to ensure reliable energy supply for the whole system.

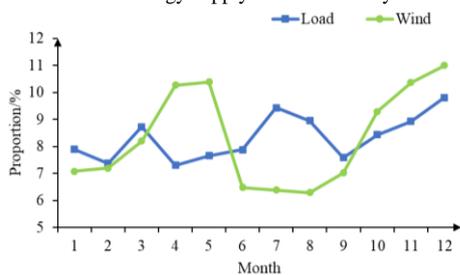

(a) The proportion curve of monthly wind power output and load.

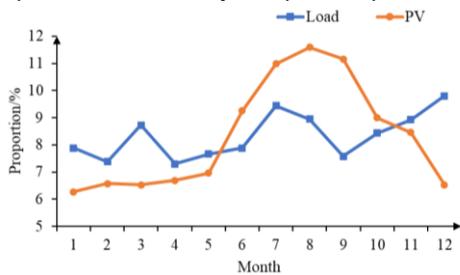

(b) The proportion curve of monthly PV power output and load.

**Fig. 1.** The proportion curve of monthly wind and PV power output and load for the whole year of a province in China.

Concentrating solar power (CSP) units have the potential to completely replace conventional thermal units by integrating the emergency boiler to cope with extreme weather [9]. It has gotten a lot of attention recently. The CSP generation climbed from $9\times10^8$ kWh in 2009 to $1.45\times10^{10}$ kWh in 2020, according to International Energy Agency statistics. By the end of 2022, the total newly built capacity of CSP units around the world will be about 7050 MW, which is 3.68% higher than 2021. In recent years, an interesting number of CSP projects under development as a result of considerable cost reductions and increased policy support in China, Morocco and South Africa [10]. It clearly proposes to accelerate the construction of high-quality peaking regulation resources in the Fourteenth Five-year Plan for energy development in China [11]. The study of advanced energy storage technology, especially CSP technology, is very important for the development of power grids according to the Action Plan for Energy Technology Revolution Innovation (2016-2030) in China [12].

Many researchers have studied the model and configuration of CSP plants [13-16]. For example, S. M. Flueckiger et al. developed a static energy flow model to analyze the mechanism of the CSP [13]. F. Zaversky et al. have proposed a transient model of the thermal energy storage (TES) device in CSP, and simulated the operation of the system considering typical weather conditions and location of CSP station [14]. Reference [15] studied the optimal design methodology of CSP during the year. A combined configuration with concentrating solar power tower and parabolic trough is proposed in reference [16] to achieve a more stable power profile. Several works study the optimal trading strategy for CSP in markets. Reference [17] has designed a profit-sharing mechanism among wind and CSP producers to effectively mitigate the real-time imbalance of wind power. Research in [18] examines the competitiveness of CSP-biomass hybrid facilities in the day-ahead market. Reference [19] investigates the CSP plant's optimal risk-based offering curve considering uncertainties in market prices and solar irradiation. The optimal offering strategy of CSP considering it provides both energy in the energy market and reserve and regulation service in the ancillary market which enhances the flexibility and economic viability of the system is proposed in reference [20-21]. Moreover, a three-stage stochastic bi-level model is employed in [22] to examine the influence of CSP in markets.

Some studies focus on the optimal behaviors and potential value of CSP from the standpoint of energy system operation and planning. Reference [23] demonstrated that CSP units with storage facilities play a buffer role towards a fully renewable penetrated energy system as their high utilization factors and competitive costs. Research in [24] modified the scheduling model of power system by incorporating CSP units, and the results show that CSP can lower the operating costs of the system and accommodate more renewable generation. Furthermore, Gökçe Kahvecioğlu et al. developed a dispatch optimization method of a combined CSP system considering the price and solar irradiance uncertainty simultaneously [25]. Reference [26-31] verified the value of complementary effects of wind and CSP on power system operation. For instance, a coordinated optimization method for wind and CSP units based on a multi-stage robust scheduling model is proposed in [26]. The results show that it decreased the scheduling cost by 17.22% and renewable curtailment by 57.39%. In reference [32-33], the impact of CSP capacity, storage size, and solar resources on the optimal combination of CSP and PV is analyzed. Reference [34] studied the short-term operation of power systems integrating CSP under high renewable penetration and estimate the value of CSP in minimizing renewable curtailment. A capacity expansion planning model of power system integrating CSP is established in [35], which showed the function of CSP to provide flexibility of the system. In reference [36], a coordinated expansion planning method of the power system with CSP which considers generating units, transmission lines and demand side resources together indicates exceptional peak regulation performance at a low overall cost. Additionally, a coordinated panning model considering joint power supply by wind, thermal and CSP units is established in [37], in which CSP is considered to replace some traditional thermal units to realize peak regulation function. Research in [38] explored how to transfer towards high renewable energy penetrations with the contribution of operational dispatchability of CSP. Study in [39] demonstrates that combination of CSP and CHP is a promising method to achieve a renewable-dominated energy system for system planning in the future.

However, few of them consider the potential benefits of CSP in the long-term system expansion planning for seasonal energy balance, which merits more research.

The key contribution of this paper are as follows:
- Firstly, based on the output characteristics of CSP plants and system hourly energy demand, we analyze the ability of CSP plants to replace traditional thermal units to cope with seasonal energy imbalance of the energy system under high renewable penetration.
- Secondly, an expansion planning model of energy systems integrating CSP and electric boilers (EB) under high renewable penetration which considers the investment and operation together in full hourly resolution for the whole year based on a fast cluster optimization method is proposed.
- Finally, we explore the complementary advantages of CSP in joint power and heat supply with CHP under the "Clean Heating" background and provide a comprehensive tech-economic assessment of the energy system under high renewable penetration which integrates CSP as a seasonal peak regulation source.

The paper includes four additional sections. Section II analyzes the ability of CSP to cope with seasonal energy imbalance. Section III describes the problem formulation. In section IV, the case study and discussions are presented and the conclusions are given in Section V.

## 2. Analysis of the ability of concentrating solar power plants to cope with seasonal energy imbalance

### 2.1. Analysis of the operational characteristics of concentrating solar power plants

The CSP plant usually consists of solar field (SF), TES device, and power block (PB) [40] as shown in Fig. 2. It can convert the solar energy into thermal energy by SF, transfer thermal energy by heat-transfer fluid (HTF), store thermal energy by TES which makes CSP have a stronger regulating ability, and convert thermal energy into electrical energy by the steam turbine in PB which makes CSP have a faster adjustment rate than traditional thermal units.

The schematic diagram of peak regulation of CSP is presented in Fig. 3. The peak-valley gap of the equivalent load due to wind and PV power output is much wider than the actual load of the system, thus it requires more flexible peak regulation source for the energy system under high renewable penetration. The CSP plant can increase its power output by TES discharging to make up for energy shortage, and decrease its power output by TES charging to consume extra wind and PV power.

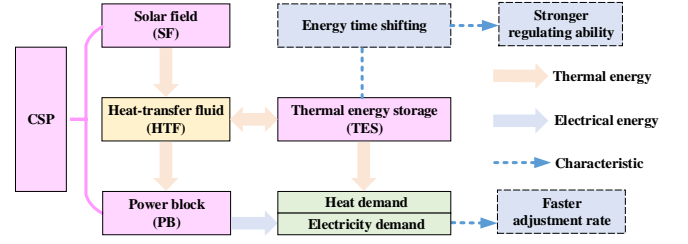

**Fig. 2.** The structure and characteristics of CSP.

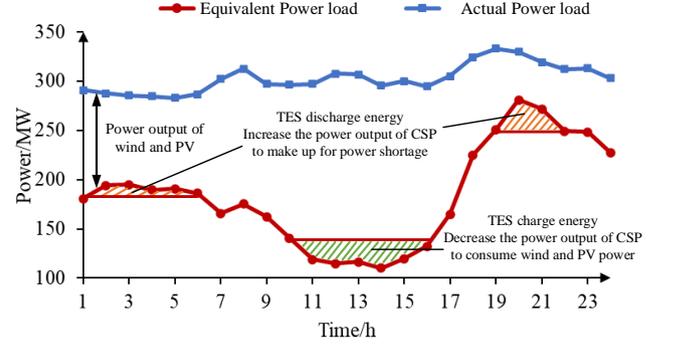

**Fig. 3.** The schematic diagram of peak regulation of CSP.

However, the CSP will also be affected by solar resources because it is a renewable energy generation technology that produces very little carbon emission. During continuous extreme weather days, the output of CSP may be reduced. But we can integrate the emergency boilers for energy shortage to cope with extreme weather.

The emergency reserve power output required by CSP $\Delta P_t^{CSP}$ to cope with electrical energy shortage $\Delta P_t$ can be calculated as:

1) If $P_t^{CSP} < I^{CSP}$, $\Delta P_t \geq I^{CSP} - P_t^{CSP}$,
Then $\Delta P_t^{CSP} = I^{CSP} - P_t^{CSP}$
2) If $P_t^{CSP} < I^{CSP}$, $\Delta P_t < I^{CSP} - P_t^{CSP}$,
Then $\Delta P_t^{CSP} = \Delta P_t$
3) If $P_t^{csp} = I^{csp}$,
Then $\Delta P_t^{CSP} = 0$

Therefore, CSP is a potentially effective technology to replace traditional thermal units by integrating emergency boiler to cope with extreme weather, and can meet long-time energy balance needs as a seasonal peak regulation source under the background of gradual compression of the traditional thermal units construction and the promotion of "Clean Heating" policy.

### 2.2. Improved CSP model based on fast cluster optimization method

Incorporating hourly unit commitment model into the planning model of energy system under high renewable penetration is widely considered. However, it poses significant challenges to integrate 8760-hour operation simulation of the large-scale energy systems on annual basis for long-term planning because it is computationally expensive and time-consuming. To address this issue, we introduce an improved CSP model based on fast cluster optimization method into the high renewable penetrated energy system planning for seasonal energy balance.

*1) Traditional CSP model*

$$P_{i,\min}^{CSP} \cdot x_{i,t}^{CSP} \leq P_{i,t}^{CSP} \leq P_{i,\max}^{CSP} \cdot x_{i,t}^{CSP} \quad (1)$$

$$P_{i,\min}^{CSP} = \underline{\alpha}_{i,t}^{CSP} \cdot P_{i,N}^{CSP} \quad (2)$$

$$P_{i,\max}^{CSP} = \overline{\alpha}_{i,t}^{CSP} \cdot P_{i,N}^{CSP} \quad (3)$$

$$-\Delta P_{i,down}^{CSP} \leq P_{i,t}^{CSP} - P_{i,t-1}^{CSP} \leq \Delta P_{i,up}^{CSP} \quad (4)$$

$$(x_{i,t}^{CSP} - x_{i,t-1}^{CSP})T_{i,on}^{CSP} + \sum_{k=t-T_{on}}^{t-1} x_{i,k}^{CSP} \geq 0 \quad (5)$$

$$(x_{i,t-1}^{CSP} - x_{i,t}^{CSP})T_{i,off}^{CSP} + \sum_{k=t-T_{off}}^{t-1} (1 - x_{i,k}^{CSP}) \geq 0 \quad (6)$$

$$Q_{i,t}^{SF-HTF} + Q_{i,t}^{TES-HTF} = Q_{i,t}^{HTF-TES} + Q_{i,t}^{HTF-PB} \quad (7)$$

$$Q_{i,t}^{SF-HTF} = \eta_{SF} \cdot S_{SF} \cdot DNI - Q_{i,t}^{cur} \quad (8)$$

$$Q_{i,t}^{CSP} = (1-\gamma \cdot \Delta t) \cdot Q_{i,t-1}^{CSP} + (Q_{i,t}^{TES,cha} - Q_{i,t}^{TES,dis}) \cdot \Delta t \quad (9)$$

$$Q_{i,t}^{TES,cha} = \eta_{TES}^{cha} \cdot (Q_{i,t}^{HTF-TES} + Q_{n,t}^{CHP,cur} + Q_{t}^{EB-TES}) \quad (10)$$

$$Q_{i,t}^{TES,dis} = (Q_{i,t}^{TES-HTF} + Q_{i,t}^{TES-HD}) / \eta_{TES}^{dis} \quad (11)$$

$$Q_{i,t}^{TES,cha} \cdot Q_{i,t}^{TES,dis} = 0 \quad (12)$$

$$Q_{i,t=0}^{TES} = Q_{i,t=T}^{TES} \quad (13)$$

$$Q_{i,\min}^{CSP} \leq Q_{i,t}^{CSP} \leq Q_{i,\max}^{CSP} \quad (14)$$

$$Q_{i,t}^{HTF-PB} = P_{i,t}^{CSP} / \eta_{PB} \quad (15)$$

*2) Fast cluster optimization method*

As for the long-term planning of the high renewable penetrated energy system on 8760-hour simulation basis, it is not feasible computationally to incorporate the traditional CSP model.

Firstly, we cluster individual units with comparable operational characteristics in neighboring places and establish the model by three integer variables $\hat{S}_{CSP,j,t}^{U}$, $\hat{S}_{CSP,j,t}^{D}$ and $\hat{S}_{CSP,j,t}^{O}$ which represent the sum of start-up capacity, shut-down capacity and online capacity of CSP group respectively, thus we can reduce the tremendous number of decision variables in the traditional model.

$$\hat{S}_{CSP,j,t}^{O} = \sum_{i=1}^{I} \left(x_{i,t}^{CSP} \cdot P_{i,N}^{CSP}\right), \quad x_{i,t}^{CSP} \in \{0,1\}, \quad i=1,...,I \quad (16)$$

$$\hat{S}_{CSP,j,t}^{U} = \sum_{i=1}^{I} \left(u_{i,t}^{CSP} \cdot P_{i,N}^{CSP}\right), \quad u_{i,t}^{CSP} \in \{0,1\}, \quad i=1,...,I \quad (17)$$

$$\hat{S}_{CSP,j,t}^{D} = \sum_{i=1}^{I} \left(d_{i,t}^{CSP} \cdot P_{i,N}^{CSP}\right), \quad d_{i,t}^{CSP} \in \{0,1\}, \quad i=1,...,I \quad (18)$$

Then, we introduce three continuous variables $S_{CSP,j,t}^{U}$, $S_{CSP,j,t}^{D}$ and $S_{CSP,j,t}^{O}$ to approximate the integer variables $\hat{S}_{CSP,j,t}^{U}$, $\hat{S}_{CSP,j,t}^{D}$ and $\hat{S}_{CSP,j,t}^{O}$ respectively to linearize the model, and they satisfy:

$$0 \leq S_{CSP,j,t}^{O}, S_{CSP,j,t}^{U}, S_{CSP,j,t}^{D} \leq S_{CSP,j} \quad (19)$$

$$S_{CSP,j} = \sum_{i=1}^{I} P_{i,N}^{CSP} \quad (20)$$

$$S_{CSP,j,t}^{O} - S_{CSP,j,t-1}^{O} = S_{CSP,j,t}^{U} - S_{CSP,j,t}^{D} \quad (21)$$

The relationship between these continuous variables is shown in Fig. 4.

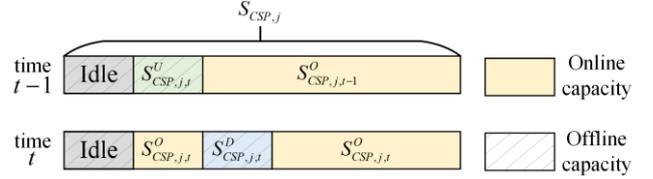

**Fig. 4.** The relationship between introduced continuous variables.

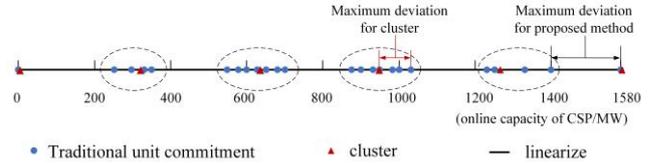

**Fig. 5.** The maximum deviation for fast cluster optimization method.

It can be seen from Fig. 5 that the maximum difference of fast cluster optimization method for total online capacity $S_{CSP,j,t}^{O}$ will not be exceed the capacity of the largest unit within the CSP group, and it will greatly fall as the number of units within the group grows.

By implementing the fast cluster optimization method [41] into the traditional CSP model, we can derive an improved CSP model which maintain accuracy and substantially fast the computational speed at the same time. Therefore, it is appropriate for the long-term planning of the energy system under high renewable penetration which considers CSP for seasonal energy balance.

*3) Improved CSP model*

$$P_{j,\min}^{CSP} \leq P_{j,t}^{CSP} \leq P_{j,\max}^{CSP} \quad (22)$$

$$P_{j,\min}^{CSP} = \underline{A}_{j,t}^{CSP} \cdot S_{CSP,j,t}^{O} \quad (23)$$

$$P_{j,\max}^{CSP} = \overline{A}_{j,t}^{CSP} \cdot S_{CSP,j,t}^{O} \quad (24)$$

$$\underline{A}_{j,t}^{CSP} = \sum_{i=1}^{J} \left(\underline{\alpha}_{i,t}^{CSP} \cdot P_{i,N}^{CSP}\right) / S_{CSP,j} \quad (25)$$

$$\overline{A}_{j,t}^{CSP} = \sum_{i=1}^{J} \left(\overline{\alpha}_{i,t}^{CSP} \cdot P_{i,N}^{CSP}\right) / S_{CSP,j} \quad (26)$$

$$\begin{aligned}P_{j,t}^{CSP} - P_{j,t-1}^{CSP} \geq{} & \underline{A}_{j,t}^{CSP} \cdot S_{CSP,j,t}^{U} - \underline{A}_{j,t}^{CSP} \cdot S_{CSP,j,t}^{D} \\ & - R_{CSP,j}^{D}\left(S_{CSP,j,t}^{O} - S_{CSP,j,t}^{U} - S_{CSP,j,t-1}^{U}\right)\end{aligned} \quad (27)$$

$$\begin{aligned}P_{j,t}^{CSP} - P_{j,t-1}^{CSP} \leq{} & \underline{A}_{j,t}^{CSP} \cdot S_{CSP,j,t}^{U} - \underline{A}_{j,t}^{CSP} \cdot S_{CSP,j,t}^{D} \\ & + R_{CSP,j}^{U}\left(S_{CSP,j,t}^{O} - S_{CSP,j,t}^{U} - S_{CSP,j,t+1}^{D}\right)\end{aligned} \quad (28)$$

$$\begin{aligned}P_{j,t}^{CSP} \leq{} & \overline{A}_{j,t}^{CSP} \cdot \left(S_{CSP,j,t}^{O} - S_{CSP,j,t}^{U} - S_{CSP,j,t+1}^{D}\right) \\ & + \underline{A}_{j,t}^{CSP} \cdot S_{CSP,j,t}^{U} + \underline{A}_{j,t}^{CSP} \cdot S_{CSP,j,t+1}^{D}\end{aligned} \quad (29)$$

$$0 \leq S_{CSP,j,1}^{D} \leq S_{CSP,j,0}^{O} \quad (30)$$

$$0 \leq S_{CSP,j,t+1}^{D} \leq S_{CSP,j,t}^{O} - \sum_{\tau=0}^{t-1} S_{CSP,j,t-\tau}^{U},$$
$$1 \leq t \leq T_{CSP,j}^{U} - 1 \quad (31)$$

$$0 \leq S^D_{CSP,j,t+1} \leq S^O_{CSP,j,t} - \sum_{\tau=0}^{T^U_{CSP,j}-2} S^U_{CSP,j,t-\tau}, \quad (32)$$
$$T^U_{CSP,j} \leq t \leq T-1$$

$$0 \leq S^U_{CSP,j,1} \leq S_{CSP,j} - S^O_{CSP,j,0} \quad (33)$$

$$0 \leq S^U_{CSP,j,t+1} \leq S^{CSP}_j - S^O_{CSP,j,t} - \sum_{\tau=0}^{t-1} S^D_{CSP,j,t-\tau}, \quad (34)$$
$$1 \leq t \leq T^D_{CSP,j}-1$$

$$0 \leq S^U_{CSP,j,t+1} \leq S_{CSP,j} - S^O_{CSP,j,t} - \sum_{\tau=0}^{T^U_{CSP,j}-2} S^D_{CSP,j,t-\tau}, \quad (35)$$
$$T^D_{CSP,j} \leq t \leq T-1$$

$$Q^{SF-HTF}_{j,t} + Q^{TES-HTF}_{j,t} = Q^{HTF-TES}_{j,t} + Q^{HTF-PB}_{j,t} \quad (36)$$

$$Q^{SF-HTF}_{j,t} = \eta_{SF} \cdot S_{SF} \cdot DNI - Q^{cur}_{j,t} \quad (37)$$

$$Q^{CSP}_{j,t} = (1-\gamma \cdot \Delta t) \cdot Q^{CSP}_{j,t-1} + (Q^{TES,cha}_{j,t} - Q^{TES,dis}_{j,t}) \cdot \Delta t \quad (38)$$

$$Q^{TES,cha}_{j,t} = \eta^{cha}_{TES} \cdot (Q^{HTF-TES}_{j,t} + Q^{CHP,cur}_{h,t} + Q^{EB-TES}_t) \quad (39)$$

$$Q^{TES,dis}_{j,t} = (Q^{TES-HTF}_{j,t} + Q^{TES-HD}_{j,t})/\eta^{dis}_{TES} \quad (40)$$

$$Q^{TES,cha}_{j,t} \cdot Q^{TES,dis}_{j,t} = 0 \quad (41)$$

$$Q^{TES}_{j,t=0} = Q^{TES}_{j,t=T} \quad (42)$$

$$Q^{CSP}_{j,\min} \leq Q^{CSP}_{j,t} \leq Q^{CSP}_{j,\max} \quad (43)$$

$$Q^{HTF-PB}_{j,t} = P^{CSP}_{j,t}/\eta_{PB} \quad (44)$$

## 3. Problem formulation

### 3.1. Framework

The seasonal energy imbalance happens in the long-term timescale. However, the traditional planning model usually simulate the operation within the typical days, which will lose the seasonal energy fluctuation characteristics. To address this issue, we propose a high-resolution planning model based on fast cluster optimization method within 8760 hours for a whole year, which is convenient for the analysis of seasonal energy imbalance problem. The framework of the energy system planning considering CSP for seasonal energy balance under high renewable penetration is shown in Fig. 6. By inputting the nameplate capacity, operational parameters and cost factor of all kinds of units which include traditional coal, wind, photovoltaic (PV), combined heat and power (CHP), CSP units and EB, the energy demand which include electricity demand and hear demand, and other information include wind and solar resource data and low carbon policy, we can obtain the monthly energy balance results, hourly energy dispatch curves, associated costs of all kind of units, renewable curtailment, carbon emissions and so on.

In the prosed model, the investment decisions and operational decisions are interlinked and optimized simultaneously. The decision variables on the investment side include newly build capacity for traditional coal unit, wind unit, PV unit, CHP unit, CSP unit and EB. The decision variables on the operational side include hourly power output of traditional coal unit, wind unit, PV unit, CHP unit and CSP unit, hourly heat output of CHP unit, CSP unit and EB, charging/discharging energy and stored energy levels for TES of CSP and operating status for all kinds of unit.

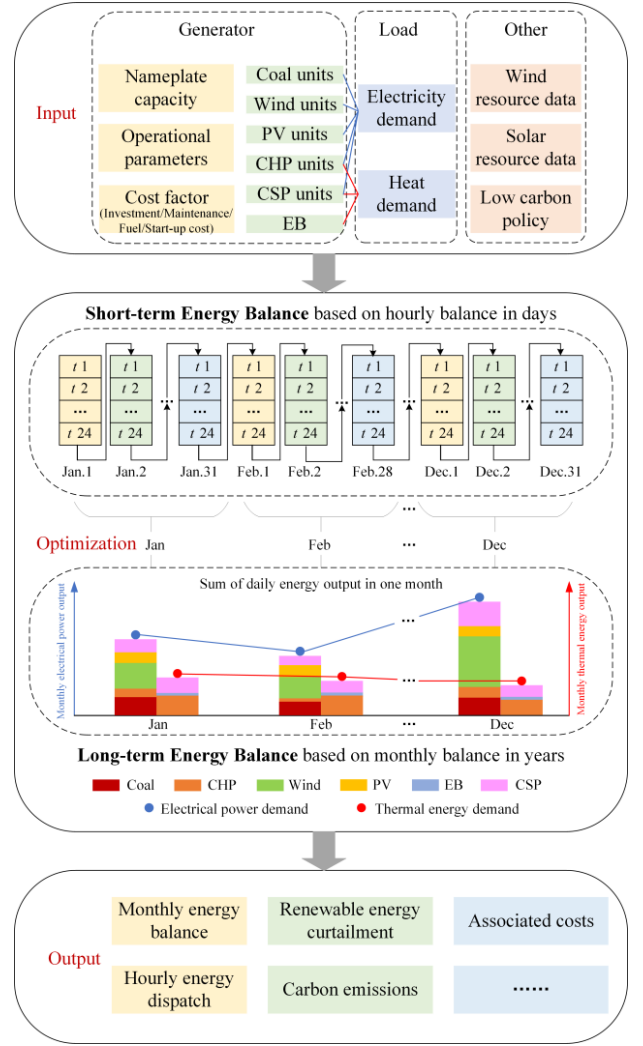

**Fig. 6.** The framework of the energy system planning considering CSP for seasonal energy balance under high renewable penetration.

### 3.2. Objective function

$$\min C = C_{coal} + C_w + C_s + C_{CSP} + C_{CHP} + C_{EB} + C_c \quad (45)$$

$$C_{coal} = \sum_{m=1}^{M} a_{coal,m} \cdot I_{coal,m} + \sum_{m=1}^{M} f_{coal,m} \cdot \bar{I}_{coal,m}$$
$$+ \sum_{m=1}^{M} \sum_{t=1}^{T} c_{coal,m} \cdot P^{coal}_{m,t} \cdot \Delta t \quad (46)$$
$$+ \sum_{m=1}^{M} \sum_{t=1}^{T} st_{coal,m} \cdot S^U_{coal,m,t}$$

$$C_w = a_w \cdot I_w + f_w \cdot \bar{I}_w \quad (47)$$

$$C_s = a_s \cdot I_s + f_s \cdot \bar{I}_s \quad (48)$$

$$C_{CSP} = \sum_{j=1}^{J} a_{CSP,j} \cdot I_{CSP,j} + \sum_{j=1}^{J} f_{CSP,j} \cdot \bar{I}_{CSP,j} \quad (49)$$

$$C_{CHP} = \sum_{h=1}^{H} a_{CHP,h} \cdot I_{CHP,h} + \sum_{h=1}^{H} f_{CHP,h} \cdot \bar{I}_{CHP,h}$$
$$+ \sum_{h=1}^{H}\sum_{t=1}^{T} c_{CHP,h} \cdot (P_{h,t}^{CHP} + c_{v,h}Q_{h,t}^{CHP}) \cdot \Delta t \quad (50)$$
$$+ \sum_{h=1}^{H}\sum_{t=1}^{T} st_{CHP,h} \cdot S_{CHP,h}^{U}$$

$$C_{EB} = a_{EB} \cdot I_{EB} + f_{EB} \cdot \bar{I}_{EB} + \sum_{t=1}^{T} c_{EB} \cdot P_{t}^{EB} \quad (51)$$

$$C_c = \sum_{t=1}^{T} c_c \cdot \left(P_{t,\max}^{w} - P_t^{w}\right) + \sum_{t=1}^{T} c_c \cdot \left(P_{t,\max}^{s} - P_t^{s}\right) \quad (52)$$

### 3.3. Investment constraints

$$0 \leq P_{m,t}^{coal} \leq \bar{P}_{m,t}^{coal} \leq \bar{I}_{coal,m} = I_{coal,m}^{0} + I_{coal,m} \quad (53)$$
$$0 \leq P_t^{w} \leq \alpha_t \cdot \bar{I}_w = \alpha_t \cdot \left(I_w^0 + I_w\right) \quad (54)$$
$$0 \leq P_t^{s} \leq \beta_t \cdot \bar{I}_s = \beta_t \cdot \left(I_s^0 + I_s\right) \quad (55)$$
$$0 \leq P_{j,t}^{CSP} \leq \lambda_t \cdot \bar{I}_{CSP,j} = \lambda_t \cdot \left(I_{CSP,j}^{0} + I_{CSP,j}\right) \quad (56)$$
$$0 \leq P_{h,t}^{CHP} \leq \bar{P}_{h,t}^{CHP} \leq \bar{I}_{CHP,h} = I_{CHP,h}^{0} + I_{CHP,h} \quad (57)$$
$$0 \leq P_t^{EB} \leq \bar{I}_{EB} = I_{EB}^0 + I_{EB} \quad (58)$$

### 3.4. Operation constraints

#### 1) Power balance constraint

$$\sum_{m=1}^{M} P_{m,t}^{coal} + P_t^{w} + P_t^{s} + \sum_{j=1}^{J} P_{j,t}^{CSP} + \sum_{h=1}^{H} P_{h,t}^{CHP} = D_{E,t} + P_t^{EB} \quad (59)$$

#### 2) Heat balance constraint

$$\sum_{j=1}^{J} Q_{j,t}^{TES,dis} + \sum_{h=1}^{H} (Q_{h,t}^{CHP} - Q_{h,t}^{CHP,cur}) + Q_t^{EB} = D_{H,t} \quad (60)$$

#### 3) System reserve constraint

$$\sum_{m=1}^{M} \bar{\mu}_{coal,m} \cdot S_{coal,m,t}^{O} + \alpha_t \cdot \bar{I}_w + \beta_t \cdot \bar{I}_s$$
$$+ \lambda_t \cdot \sum_{j=1}^{J} \bar{I}_{CSP,j} + \sum_{h=1}^{H} \bar{\mu}_{CHP,h} \cdot S_{CHP,h,t}^{O} \quad (61)$$
$$\geq D_{E,t} + R_t^d + R_w \cdot P_t^{w} + R_s \cdot P_t^{s} + R_c \cdot \sum_{j=1}^{J} P_{j,t}^{CSP}$$

#### 4) Low-carbon policy constraint

$$P_t^{w} + P_t^{s} + \sum_{j=1}^{J} P_{j,t}^{CSP} \geq p \cdot D_{E,t} \quad (62)$$

#### 5) CSP constraint

The CSP constraints are shown in equation (22)-(44).

#### 6) CHP constraint

$$Q_{h,\min}^{CHP} \leq Q_{h,t}^{CHP} \leq Q_{h,\max}^{CHP} \quad (63)$$

$$P_{h,t}^{CHP} \geq \max\{c_{m,h}Q_{h,t}^{CHP} - (c_{m,h}+c_{v,h})Q_{h,\max}^{CHP} + P_{h,\max}^{CHP},$$
$$P_{h,\min}^{CHP} - c_{v,h}Q_{h,t}^{CHP}\} \quad (64)$$

$$P_{h,t}^{CHP} \leq P_{h,\max}^{CHP} - c_{v,h}Q_{h,t}^{CHP} \quad (65)$$

$$(P_{h,t}^{CHP} + c_{v,h}Q_{h,t}^{CHP}) - (P_{h,t-1}^{CHP} + c_{v,h}Q_{h,t-1}^{CHP})$$
$$\geq \underline{A}_{h,t}^{CHP} \cdot S_{CHP,h,t}^{U} - \underline{A}_{h,t}^{CHP} \cdot S_{CHP,h,t}^{D} \quad (66)$$
$$- R_{CHP,h}^{D} \left(S_{CHP,h,t}^{O} - S_{CHP,h,t}^{U} - S_{CHP,h,t-1}^{U}\right)$$

$$(P_{h,t}^{CHP} + c_{v,h}Q_{h,t}^{CHP}) - (P_{h,t-1}^{CHP} + c_{v,h}Q_{h,t-1}^{CHP})$$
$$\leq \underline{A}_{h,t}^{CHP} \cdot S_{CHP,h,t}^{U} - \underline{A}_{h,t}^{CHP} \cdot S_{CHP,h,t}^{D} \quad (67)$$
$$+ R_{CHP,h}^{U} \left(S_{CHP,h,t}^{O} - S_{CHP,h,t}^{U} - S_{CHP,h,t+1}^{D}\right)$$

$$P_{h,t}^{CHP} + c_{v,h}Q_{h,t}^{CHP}$$
$$\leq \bar{A}_{h,t}^{CHP} \cdot \left(S_{CHP,h,t}^{O} - S_{CHP,h,t}^{U} - S_{CHP,h,t+1}^{D}\right) \quad (68)$$
$$+ \underline{A}_{h,t}^{CHP} \cdot S_{CHP,h,t}^{U} + \underline{A}_{h,t}^{CHP} \cdot S_{CHP,h,t+1}^{D}$$

$$0 \leq S_{CHP,h,t}^{O} \leq S_{CHP,h} \quad (69)$$

$$S_{CHP,h,t}^{O} - S_{CHP,h,t-1}^{O} = S_{CHP,h,t}^{U} - S_{CHP,h,t}^{D} \quad (70)$$

$$S_{CHP,h} = \sum_{i=1}^{I'} (P_{t,\max}^{CHP} + c_{v,i}Q_{i,\max}^{CHP}) \quad (71)$$

#### 7) EB constraint

$$\eta_{EB} \cdot P_t^{EB} = Q_t^{EB-TES} + Q_t^{EB} \quad (72)$$
$$Q_{\min}^{EB} \leq Q_t^{EB} \leq Q_{\max}^{EB} \quad (73)$$

### 3.5. The tech-economic assessment index

#### 1) Renewable energy curtailment

$$\eta_{cur} = \frac{\sum_{t=1}^{T}[(P_{t,\max}^{w} - P_t^{w}) + (P_{t,\max}^{s} - P_t^{s})]}{\sum_{t=1}^{T}(P_{t,\max}^{w} + P_{t,\max}^{s})} \quad (74)$$

#### 2) Carbon emissions

$$E_{CO_2,t} = \omega_1 \sum_{m=1}^{M} P_{m,t}^{coal} + \omega_2 \sum_{h=1}^{H} (P_{h,t}^{CHP} + c_{v,h}Q_{h,t}^{CHP}) \quad (75)$$

*3) Levelized cost of energy (LCOE)*

$$LCOE = \frac{\sum_{t=1}^{T} \frac{C_{inve,t} + C_{main,t} + C_{oper,t}}{(1+r)^t}}{\sum_{t=1}^{T} \frac{E_t + H_t}{(1+r)^t}} \quad (76)$$

## 4. Case study

### 4.1. Data collection

It is abundant in solar energy resources in northwest area of China, with an annual average solar radiation of 1750 kWh/m² [42]. The following study focuses on the capacity expansion of the energy system considering CSP for seasonal energy balance in Xinjiang province which is a typical area for "Clean Heating" policy with the projection to 2050 when renewable energy generation will become the main power supply. The hourly capacity factors for wind and solar power in Xinjiang province are calculated according to the NASA database [43]. In Xinjiang energy system, the total capacity of traditional thermal, wind and PV units is up to 60,599.6 MW, 26,140.5 MW and 14,672.2 MW respectively in 2022. The operational characteristics for coal, CHP and CSP units are presented in TABLE I, TABLE II and TABLE III respectively. The efficiency factor of EB is 87%. The energy demand is estimated based on the actual operating data over the years with an annual growth rate of 3.5%. The installed costs of coal, CSP and CHP units are amortized to 25 years, the installed costs of wind and PV units are amortized to 15 years, and the discount ratio of investment is 7%. The price for coal and natural gas are $37.7/ton and $0.1691/m3 in this area respectively.

**Table 1 - Operational characteristics for coal units.**

| Small (10-300MW) | | Medium (300-600MW) | | Large (600-1000MW) | |
|---|---|---|---|---|---|
| Parameter | Value | Parameter | Value | Parameter | Value |
| Max. output (%) | 100 | Max. output (%) | 100 | Max. output (%) | 100 |
| Min. output (%) | 50 | Min. output (%) | 50 | Min. output (%) | 50 |
| Min. up time (h) | 8 | Min. up time (h) | 8 | Min. up time (h) | 24 |
| Min. down time (h) | 4 | Min. down time (h) | 8 | Min. down time (h) | 48 |
| Ramping limit (%/h) | 35 | Ramping limit (%/h) | 35 | Ramping limit (%/h) | 35 |
| Fuel use (g/kwh) | 345 | Fuel use (g/kwh) | 302 | Fuel use (g/kwh) | 281 |

**Table 2 - Operational characteristics for CHP units.**

| Small (10-300MW) | | Medium (300-600MW) | | Large (600-1000MW) | |
|---|---|---|---|---|---|
| Parameter | Value | Parameter | Value | Parameter | Value |
| $c_u$ | 0.75 | $c_u$ | 0.75 | $c_u$ | 0.75 |
| $c_v$ | 0.15 | $c_v$ | 0.15 | $c_v$ | 0.15 |
| Max. output (%) | 90 | Max. output (%) | 90 | Max. output (%) | 90 |
| Min. output (%) | 60 | Min. output (%) | 60 | Min. output (%) | 60 |
| Ramping limit (%/h) | 30 | Ramping limit (%/h) | 30 | Ramping limit (%/h) | 30 |
| Fuel use (g/kwh) | 370 | Fuel use (g/kwh) | 320 | Fuel use (g/kwh) | 295 |

**Table 3 - Operational characteristics for CSP units.**

| Parameter | Value | Parameter | Value |
|---|---|---|---|
| $\eta_{SF}$ (%) | 37 | Max. output (%) | 100 |
| $\eta_{TES}^{cha}$ (%) | 98 | Min. output (%) | 15 |
| $\eta_{TES}^{dis}$ (%) | 98 | Min. up time (h) | 2 |
| $\eta_{PB}$ (%) | 40 | Min. down time (h) | 2 |
| $\gamma$ (%/h) | 0.031 | Ramping limit(%/h) | 30 |

### 4.2. Scenarios setting

To verify the validity of the proposed method in this paper, we set three scenarios as shown in Table 4. Each scenario considers a one-year hourly operational simulation when optimizing the capacity expansion.

Scenario 1: Energy systems planning without CSP plants. (Base scenario)
Scenario 2: Energy systems planning with CSP plants.
Scenario 3: Energy systems planning with CSP plants integrating electric boiler (EB) as emergency boiler to cope with extreme weather and renewable curtailment which is in line with "Electric Heating" policy at the same time.

**Table 4 - Scenarios setting.**

| Scenario | With coal | With wind | With PV | With CHP | With CSP | With EB |
|---|---|---|---|---|---|---|
| 1 | √ | √ | √ | √ | × | × |
| 2 | √ | √ | √ | √ | √ | × |
| 3 | √ | √ | √ | √ | √ | √ |

## 4.3. Results and discussion

### 1) Systemic capacity structure analysis

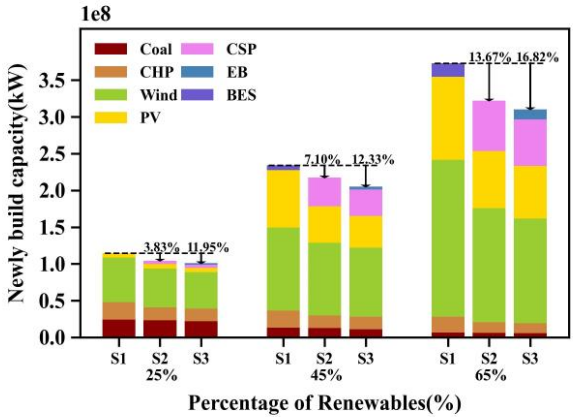

(a) Installed capacity in different scenarios.

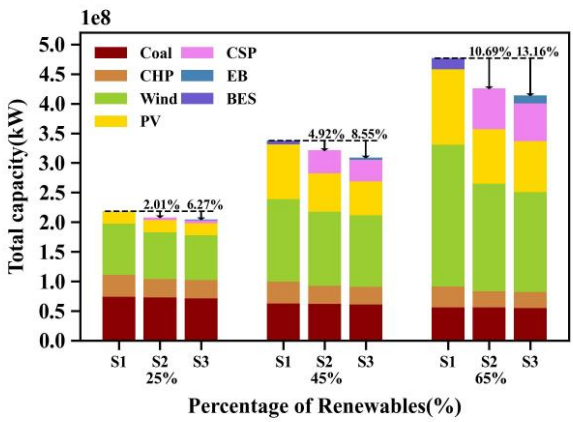

(b) Total capacity in different scenarios.

**Fig. 7.** Installed and total capacity in different scenarios.

The installed and total capacity of scenario 1, 2 and 3 which are aggregated from the investment decisions identified in the optimization model at 25%, 45% and 65% renewable penetration is compared in Fig. 7. By comparing scenario 1 and scenario 2, we can find that integrating the CSP plant can compress the installed capacity of coal, CHP, wind and PV units. The total installed capacity of all kinds of units in scenario 2 is reduced by 3.83%, 7.10%, 13.67% at 25%, 45% and 65% renewable penetration respectively compared to scenario 1. By comparing scenario 2 and scenario 3, EB as emergency boiler of CSP can further decrease the installed capacity of coal, CHP, wind and PV units. The total installed capacity of all kinds of units in scenario 3 is reduced by 11.95%, 12.33%, 16.82% at 25%, 45% and 65% renewable penetration respectively compared to scenario 1.

### 2) Generation mix analysis

The optimal generation mix of scenario 1, 2 and 3 which are derived from the hourly simulation results at 65% renewable penetration is compared in Fig. 8 and Fig. 9. The monthly energy balance of the whole year in different scenarios is shown in Fig. 8. The typical hourly energy output of different seasons in different scenarios is presented in Fig. 9.

According to Fig. 8(a), (c) and (e), the electricity demand is at its peak in summer (Jun.-Aug.), but the wind power output peaks in spring (Mar.-May) and winter (Dec.-Feb.). Thus, constrained by natural resources there is an obvious seasonal mismatch between wind power supply and electricity demand. In scenario 1, by integrating PV units with higher output in summer and lower output in winter, the seasonal imbalances can be mitigated to some extent with attribution to the complementarity of wind and solar power as shown in Fig. 8(a). However, thermal energy supply by CHP units as shown in Fig. 8(b) will reduce the system flexibility due to its limited operating characteristics, and thus there will be both high renewable curtailment rate and carbon emissions. In scenario 2, CSP units can not only participate in thermal energy supply with CHP units as shown in Fig. 8(d), but also perform as the peak regulation source to improve the operational flexibility of the system, thus it replaces a part of thermal units and decline the energy output of coal and CHP units as shown in Fig. 8(c) compared with scenario 1, which can reduce renewable energy curtailment and carbon emissions. However, as a renewable energy technology, CSP will also be limited by solar resources which may be insufficient in winter or some days with extreme weather. Therefore, in scenario 3 where EB is integrated to cope with extreme weather and renewable curtailment, the energy system will provide a more stable energy supply with less renewable curtailment while further compressing the space for traditional thermal units as shown in Fig. 8(e) and (f). Besides, we can find that the monthly electrical power input and thermal energy output of EB become more noticeable in December, January and February in winter.

According to Fig. 8(a) and Fig. 9(a), the electricity demand is satisfied by the coal, CHP, wind, PV units and BES, the peak-valley difference of the load is mainly adjusted by coal, CHP units and BES, and the peak-valley difference of the net load is $7.25 \times 10^7$ kW in scenario 1. According to Fig. 8(c) and Fig. 9(c), the electricity demand is satisfied by the coal, CHP, wind, PV and CSP units, the peak-valley difference of the load can be adjusted by coal, CHP and CSP units, and the peak-valley difference of the net load is $6.80 \times 10^7$ kW in scenario 2, which is reduced by 6.21% compared with scenario 1. Therefore, we can find that it can effectively relieve the peaking pressure of the system by integrating CSP units. It is mainly because: on the one hand, CSP units have a stronger regulating ability and a faster adjustment rate due to TES which can realize temporal shifting for energy and steam turbine in PB which can convert thermal energy into electrical power as presented in Fig. 2 and Fig. 3; on the other hand, CSP units can provide heat supply jointly with CHP units as shown in Fig. 8(d) and Fig. 9(d). It can thus effectively expand the operating range of CHP units. According to Fig. 8(e) and Fig. 9(e), the electricity demand is satisfied by the coal, CHP, wind, PV and CSP units, the peak-valley difference of the load can be adjusted by coal, CHP and CSP units, and the peak-valley difference of the net load is $5.82 \times 10^7$ kW in scenario 3, which is reduced by 14.41% compared with scenario 2, and reduced by 19.72% compared with scenario 1. It is mainly because of EB, which can provide thermal energy output with CSP and CHP together as shown in Fig. 8(f) and Fig. 9(f), further enhancing the operational flexibility of the system.

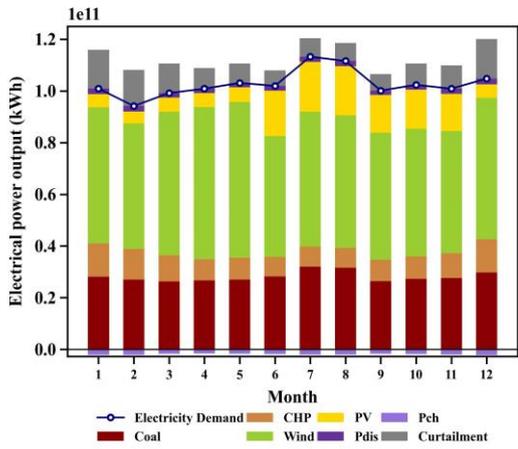
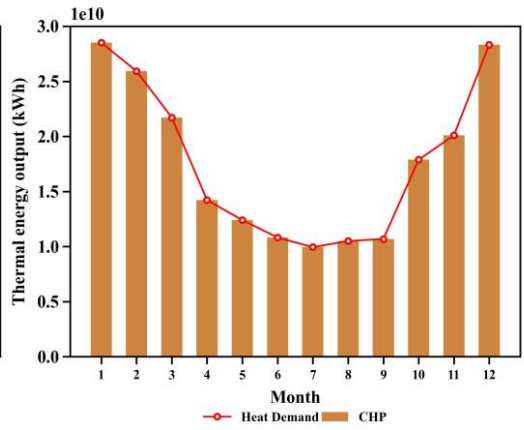

(a) Monthly electric power balance in scenario 1.　　　(b) Monthly thermal energy balance in scenario 1.

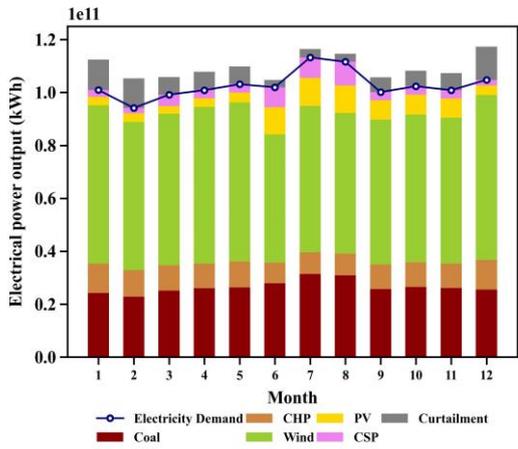
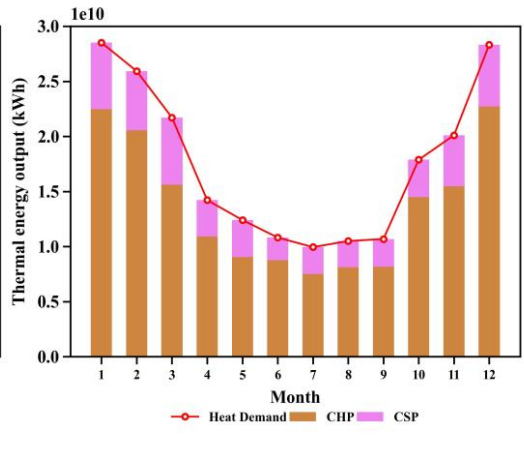

(c) Monthly electric power balance in scenario 2.　　　(d) Monthly thermal energy balance in scenario 2.

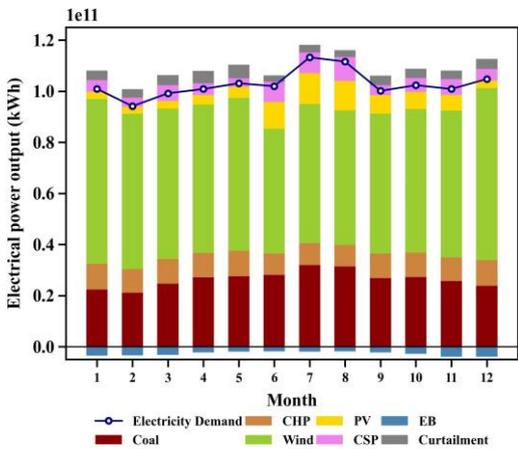
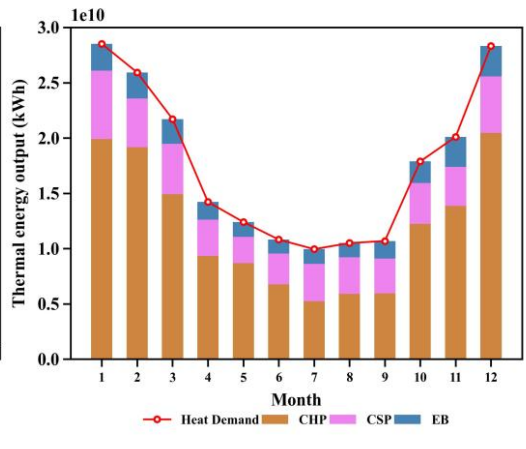

(e) Monthly electric power balance in scenario 3.　　　(f) Monthly thermal energy balance in scenario 3.

**Fig. 8. Monthly energy balance in different scenarios.**

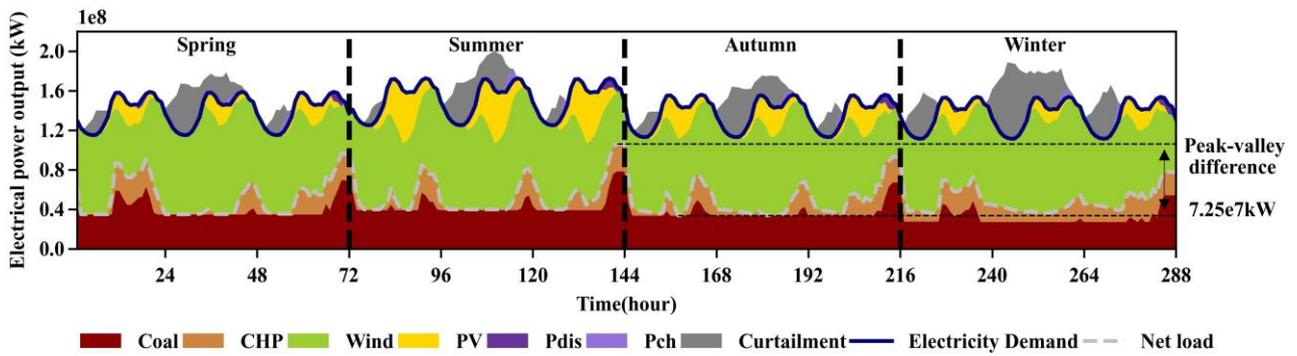

(a) Typical hourly electrical power output of different seasons in scenario 1.

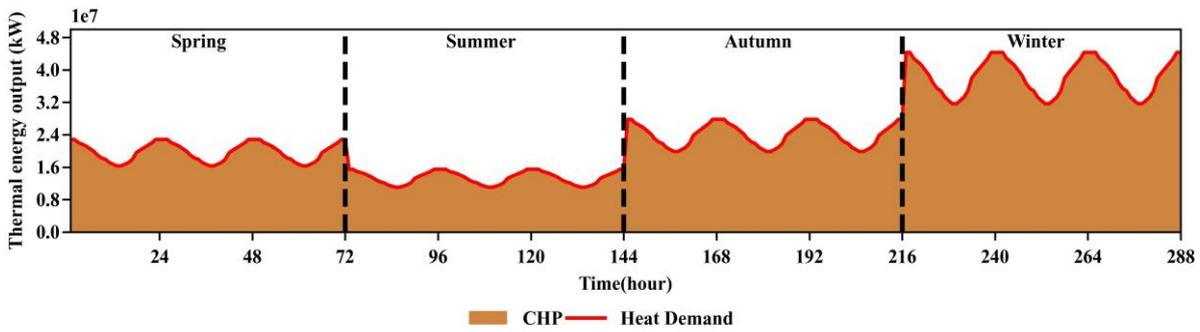

(b) Typical hourly thermal energy output of different seasons in scenario 1.

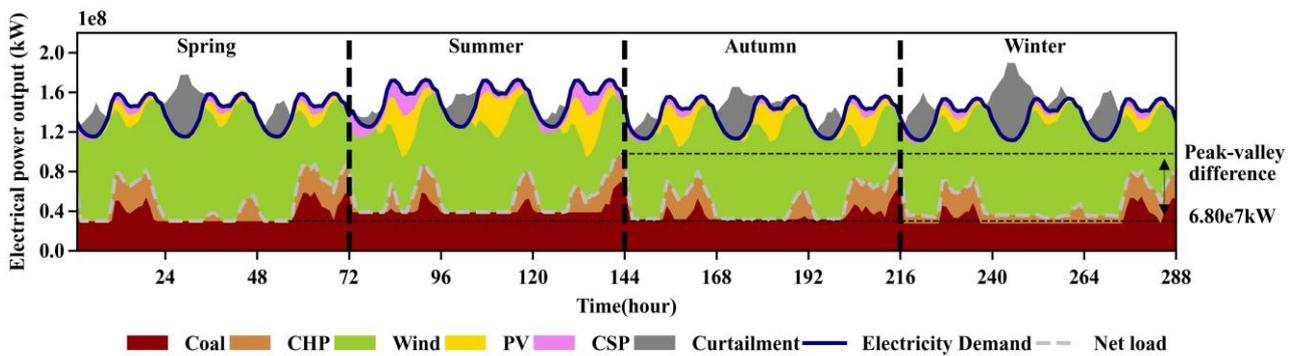

(c) Typical hourly electrical power output of different seasons in scenario 2.

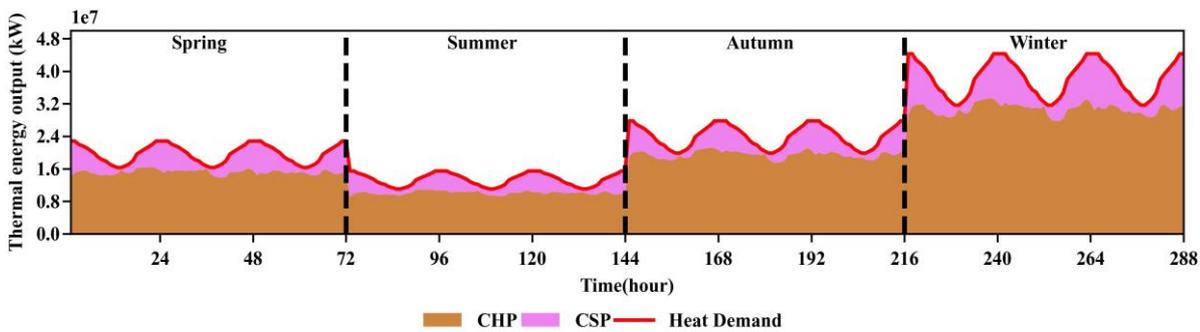

(d) Typical hourly thermal energy output of different seasons in scenario 2.

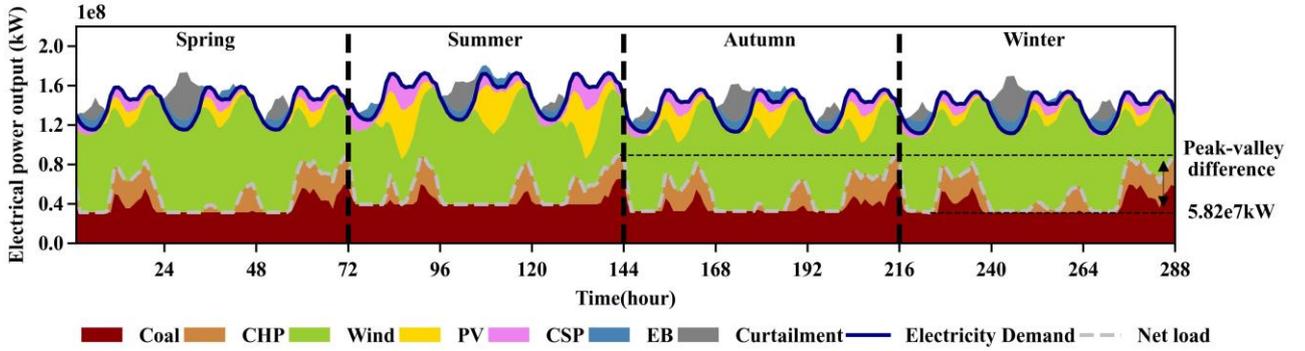

(e) Typical hourly electrical power output of different seasons in scenario 3.

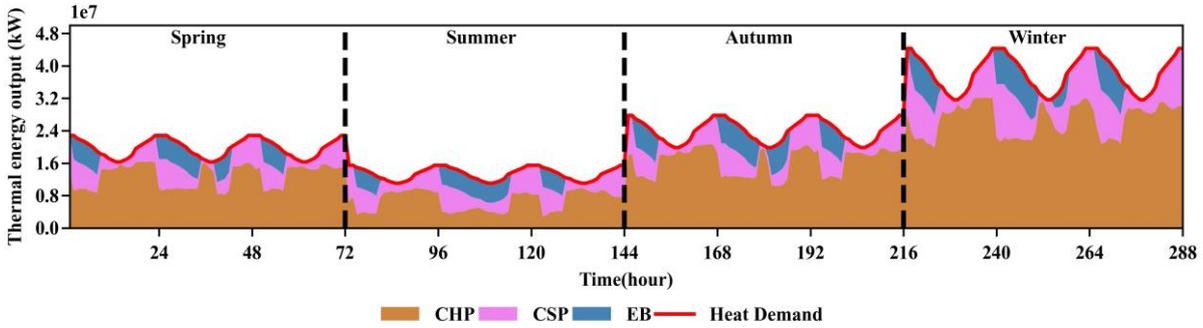

(f) Typical hourly thermal energy output of different seasons in scenario 3.
**Fig. 9.** Typical hourly energy output of different seasons in different scenarios.

### 3) Analysis of associated costs and levelized cost of energy (LCOE)

The associated cost of scenario 1, 2 and 3 at 65% renewable penetration is presented in Fig. 10. It shows that the cost of scenario 2 which introducing CSP plants is reduced by 5.86% compared to scenario 1, with contribution to the reduction of installed capacity of coal, CHP, wind and PV units as well as fuel and start-up cost of the system. The introduction of EB in scenario 3 further decreases the cost, it is reduced by 8.73% compared to scenario 1, with contribution to more compression of coal, CHP, wind and PV units and more operational flexibility provided to the energy system.

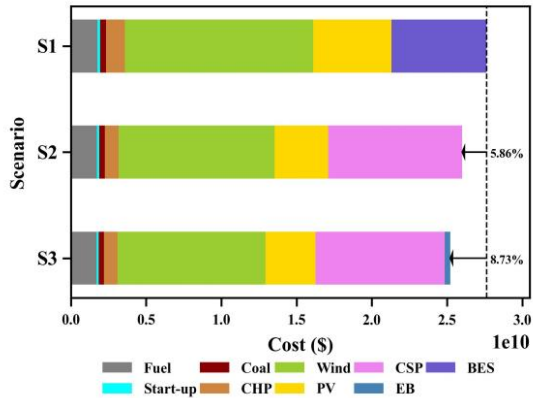

**Fig. 10.** Associated costs in different scenarios.

The levelized cost of energy (LCOE) of scenario 1, 2 and 3 at 65% renewable penetration is presented in Table 5. The LCOE of coal, CHP, wind and PV units in scenario 1 are 0.0690 $/kWh, 0.0259 $/kWh, 0.0429 $/kWh and 0.0288 $/kWh respectively. Compared with scenario 1, the LCOE of coal, CHP, wind and PV units in scenario 2 are all reduced, they are 0.0659 $/kWh, 0.0239 $/kWh, 0.0427 $/kWh and 0.0287 $/kWh respectively and the LCOE of CSP in scenario 2 is 0.0785 $/kWh. In scenario 3, the LCOE of coal, CHP, wind, PV and CSP units are all lower than scenario 2, which are 0.0651 $/kWh, 0.0227 $/kWh, 0.0426 $/kWh and 0.0286$/kWh 0.0742 $/kWh respectively. Therefore, the LOCE of all kinds units in scenario 3 are cheapest compared with scenario 1 and scenario 2.

**Table 5 - Levelized cost of energy (LCOE) in different scenarios.**

| LCOE ($/kWh) | Scenario 1 | Scenario 2 | Scenario 3 |
|---|---|---|---|
| $LCOE_{coal}$ | 0.0690 | 0.0659 | 0.0651 |
| $LCOE_{CHP}$ | 0.0259 | 0.0239 | 0.0227 |
| $LCOE_{wind}$ | 0.0429 | 0.0427 | 0.0426 |
| $LCOE_{PV}$ | 0.0288 | 0.0287 | 0.0286 |
| $LCOE_{CSP}$ | / | 0.0785 | 0.0742 |

### 4) Analysis of renewable curtailment

According to Table 6, the renewable curtailment of scenario 1, 2 and 3 is $38.71 \times 10^8$ kW, $27.35 \times 10^8$ kW and $16.85 \times 10^8$ kW at 45% renewable penetration, $94.16 \times 10^8$ kW, $71.29 \times 10^8$ kW and $43.17 \times 10^8$ kW at 65% renewable penetration and $251.41 \times 10^8$ kW, $150.25 \times 10^8$ kW and $87.21 \times 10^8$ kW at 85% renewable penetration respectively. Compared with scenario 1, the renewable energy curtailment of scenario 2 are reduced by $11.36 \times 10^8$ kW at 45% renewable penetration, $22.87 \times 10^8$ kW at 65% renewable penetration and 101.16×10⁸ kW at 85% renewable penetration respectively, the renewable energy curtailment of scenario 3 are reduced by 26.86×10⁸ kW at 45% renewable penetration, 50.99×10⁸ kW at 65% renewable penetration and 164.20×10⁸ kW at 85% renewable penetration respectively. Besides, the renewable energy curtailment rate of scenario 1 is 8.02%, 13.05% and 23.46% at 45%, 65% and 85% renewable penetration respectively. The renewable energy curtailment rate of scenario 2 is 7.16%, 11.39% and 18.30% at 45%, 65% and 85% renewable penetration respectively, which are reduced by 10.72%, 12.72% and 21.99% respectively compared with scenario 1. The renewable energy curtailment rate of scenario 3 is 3.52%, 5.45% and 8.65% at 45%, 65% and 85% renewable penetration respectively, which are reduced by 56.11%, 58.24% and 63.13% respectively compared with scenario 1. The renewable energy curtailment is noticeably reduced by flexibly peaking regulation of CSP and the consumption of abandoned wind and solar power by EB.

**Table 6 - Renewable energy curtailment in different scenarios.**

| Assessment index | Scenario | Renewable energy penetration | | |
|---|---|---|---|---|
| | | 45% | 65% | 85% |
| Renewable curtailment (×10⁸kW) | Scenario 1 | 38.71 | 94.16 | 251.41 |
| | Scenario 2 | 27.35 | 71.29 | 150.25 |
| | Scenario 3 | 16.85 | 43.17 | 87.21 |
| Reduction (×10⁸kW) | Scenario 1 | / | / | / |
| | Scenario 2 | 11.36 | 22.87 | 101.16 |
| | Scenario 3 | 26.86 | 50.99 | 164.20 |
| Renewable curtailment rate (%) | Scenario 1 | 8.02 | 13.05 | 23.46 |
| | Scenario 2 | 7.16 | 11.39 | 18.30 |
| | Scenario 3 | 3.52 | 5.45 | 8.65 |

*5) Analysis of carbon emissions*

According to Fig. 11, the carbon emissions are highest in summer due to the large amount of energy demanded in summer, and the carbon emissions in spring is similar to which in autumn. Fig. 11 also shows that we can decrease carbon emissions by integrating CSP and EB. For example, the carbon emission is 25.47 million ton in January in scenario 1, but it can be reduced to 24.99 million ton in scenario 2 which integrating CSP, and reduced to 24.70 million ton in scenario 3 which integrating CSP and EB.

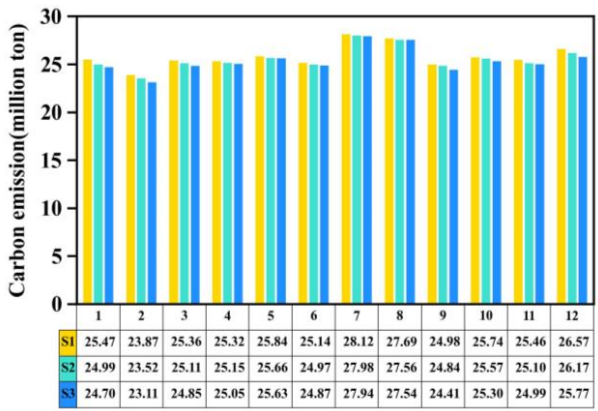

**Fig. 11.** Carbon emissions in different scenarios.

*6) Analysis of comprehensive benefit in different scenarios*

The cost, peak-valley difference of net load and renewable curtailment in scenario 1, 2 and 3 at 65% renewable penetration are summarized in Table 7. By taking these three assessment indexes as the vertices of the triangle, and enclosing the scenario 1, 2 and 3 into a corresponding triangle according to the scale respectively, we can obtain Fig. 12. It can be seen from Fig. 12 that the enclosed area of scenario 3 is the smallest. Therefore, the comprehensive benefit of scenario 3 which considering CSP and EB participating in the energy system planning is optimal.

**Table 7 - Comprehensive benefits in different scenarios.**

| Assessment index | Scenario 1 | Scenario 2 | Scenario 3 |
|---|---|---|---|
| Cost (billion $) | 27.62 | 26.01 | 25.21 |
| Peak-valley difference of net load (million kW) | 72.48 | 68.02 | 58.19 |
| Renewable curtailment rate (%) | 13.05 | 11.39 | 5.45 |

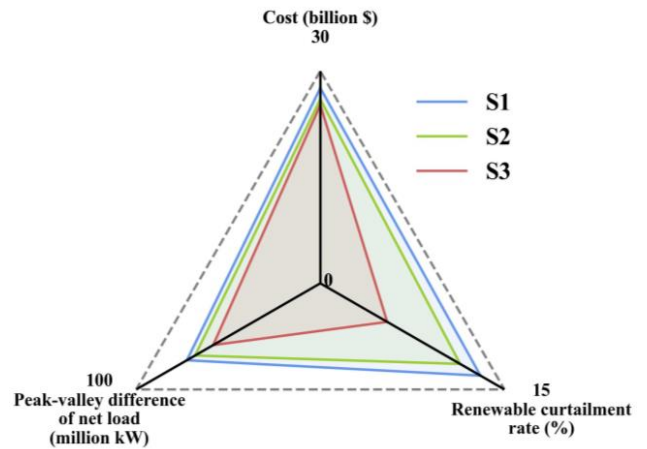

**Fig. 12.** Comprehensive benefits of 3 different scenarios.

*7) Sensitivity analysis*

**A. Impact of the thermal energy storage time of TES in CSP**

As a key component of CSP to realize the temporal shifting for energy, TES is very important for CSP to make flexible peak regulation. To study the impact of the thermal energy storage time of TES in CSP on system operation, we set four more scenarios:

Scenario 4: Energy systems planning with CSP plants in which thermal energy storage time of TES is shorter than scenario 2.

Scenario 5: Energy systems planning with CSP plants in which thermal energy storage time of TES is longer than scenario 2.

Scenario 6: Energy systems planning with CSP plants integrating EB as emergency boiler in which thermal energy storage time of TES is shorter than scenario 3.

Scenario 7: Energy systems planning with CSP plants integrating EB as emergency boiler in which thermal energy storage time of TES is longer than scenario 3.

As shown in Table 8, with the rise of thermal energy storage time of TES in CSP, the peak-valley difference rate of net load, renewable curtailment rate and carbon emission of the energy system decreases. For example, when the thermal energy storage time of TES in CSP is increased from 10 hours to 12 hours, the peak-valley difference rate is reduced from 69.14% in scenario2 to 68.27% in scenario 5, and from 67.99% in scenario 3 to 67.78% in scenario 7, the renewable curtailment rate is decreased from 11.39% in scenario 2 to 10.89% in scenario5, and from 5.45% in scenario 3 to 4.97% in scenario 7, and the carbon emission is declined from 306.62 million ton in scenario 2 to 304.96 million ton in scenario 5, and from 304.16 million ton in scenario 3 to 303.78 million ton in scenario 7. The main reason is that the longer energy storage time of TES, the more pronounced effect of temporal shifting for energy on the system, thus making the system operation more flexible.

**Table 8 – Impact of thermal energy storage time of TES in CSP on system operation.**

| scenario | Storage time of TES (h) | Peak-valley difference rate (%) | Renewable curtailment rate (%) | Carbon emission (million ton) |
|---|---|---|---|---|
| Scenario 4 | 8 | 70.38 | 12.61 | 309.24 |
| Scenario 2 | 10 | 69.14 | 11.39 | 306.62 |
| Scenario 5 | 12 | 68.27 | 10.89 | 304.96 |
| Scenario 6 | 8 | 68.24 | 6.20 | 304.71 |
| Scenario 3 | 10 | 67.99 | 5.45 | 304.16 |
| Scenario 7 | 12 | 67.78 | 4.97 | 303.78 |

### B. Impact of the investment cost of CSP

Nowadays, the expensive investment cost of CSP has a significant impact on its development. However, it has a high potential for future decrease [44]. To observe the impact of the investment cost of CSP, we set four more scenarios:

Scenario 8: Energy systems planning with CSP plants in which the investment cost of CSP is reduced to the same level as wind units compared with scenario 2.

Scenario 9: Energy systems planning with CSP plants in which the investment cost of CSP is reduced to the same level as PV units compared with scenario 2.

Scenario 10: Energy systems planning with CSP plants integrating EB as emergency boiler in which the investment cost of CSP is reduced to the same level as wind units compared with scenario 3.

Scenario 11: Energy systems planning with CSP plants integrating EB as emergency boiler in which the investment cost of CSP is reduced to the same level as PV units compared with scenario 3.

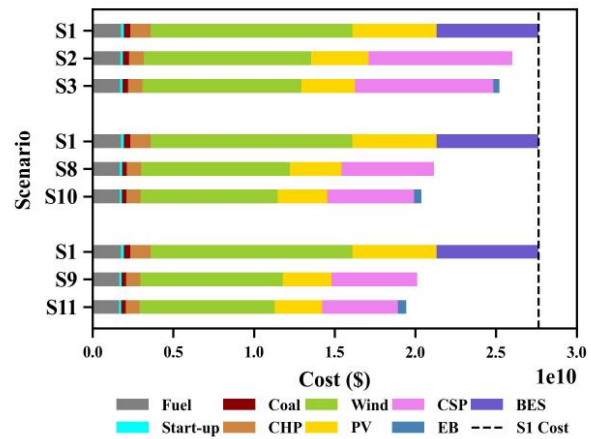

**Fig. 13.** Impact of investment cost of CSP on associated cost.

According to Fig. 13, the cost of energy system planning with CSP in scenario 8 and with CSP and EB in scenario 10 will be 23.45% and 26.26% lower than that in scenario 1 respectively when the investment cost of CSP is reduced to the same level as wind units. Moreover, the cost of energy system planning with CSP in scenario 9 and with CSP and EB in scenario 11 will be 27.20% and 29.62% lower than that in scenario 1 respectively when the investment cost of CSP is reduced to the same level as PV units, which performs well in economic.

As shown in Table 9, compared with scenario 2, with the reduction of the investment of CSP, the peak-valley difference rate of net load in scenario 8 and scenario 9 is reduced from 69.14% to 66.41% and 65.70% respectively, the renewable curtailment rate is dropped from 11.39% to 8.08% and 6.58% respectively, and the carbon emission is lessened from 306.62 to 284.41 and 276.54 million ton respectively. Compared with scenario 3, with the reduction of the investment of CSP, the peak-valley difference rate of net load in scenario 10 and scenario 11 is reduced from 67.99% to 64.52% and 63.73% respectively, the renewable curtailment rate is dropped from 5.45% to 4.14% and 3.31% respectively, and the carbon emission is lessened from 304.16 to 281.73 and 273.75 million ton respectively.

**Table 9 – Impact of investment cost of CSP on system operation.**

| scenario | Peak-valley difference rate (%) | Renewable curtailment rate (%) | Carbon emission (million ton) |
|---|---|---|---|
| Scenario 2 | 69.14 | 11.39 | 306.62 |
| Scenario 8 | 66.41 | 8.08 | 284.41 |
| Scenario 9 | 65.70 | 6.58 | 276.54 |
| Scenario 3 | 67.99 | 5.45 | 304.16 |
| Scenario 10 | 64.52 | 4.14 | 281.73 |
| Scenario 11 | 63.73 | 3.31 | 273.75 |

## 5. Conclusion

Under the background of the decarbonization pathway and "Clean Heating" promotion, the issue of seasonal imbalance in high renewable penetrated energy systems can be effectively solved by introducing CSP with EB to meet electricity and heat demand together. By applying the proposed long-term high-resolution expansion planning model in Xinjiang province in China, we can find that CSP can meet long-time energy balance as a seasonal peak regulation source and raise the utilization rate of variable renewable energy thus reduce the carbon emission helpfully with a competitive cost.

According to the comparison of different scenarios, it shows that the comprehensive benefit of scenario with CSP and EB is optimal. From the investment decision identified in the proposed model, it can reduce the total installed capacity by 11.95%, 12.33%, 16.82% at 25%, 45% and 65% renewable penetration respectively compared to the base scenario. The optimal generation mix which are derived from the hourly simulation results shows that the joint energy supply by integrating CSP and EB can reduce the cost, peak-valley difference of net load and renewable curtailment by 8.73%, 19.72% and 58.24% respectively at 65% renewable penetration compared to the base scenario.

Furthermore, by studying the impact of the thermal energy storage time of TES in CSP on system operation, we can find that with the rise of thermal energy storage time of TES, the peak-valley difference rate of net load, renewable curtailment rate and carbon emission of the energy system are further decreased. Considering the potential reduction of the investment of CSP in the future, the energy system planning with CSP and EB will perform much better in economic and operation.

## Acknowledgements

This work was supported by National Natural Science Foundation of China-Enterprise Innovation and Development Joint Fund (U22B20102).